\documentclass[runningheads]{llncs}

\usepackage{natbib}
\usepackage{graphicx}
\usepackage{amsmath}
\usepackage{amssymb}
\usepackage{booktabs}  

\begin{document}

\title{Accelerating 3D MULTIPLEX MRI Reconstruction with Deep Learning}
\author{Eric Z. Chen\inst{1} \and Yongquan Ye\inst{2} \and  Xiao Chen\inst{1} \and Jingyuan Lyu\inst{2} \and Zhongqi Zhang\inst{3} \and Yichen Hu\inst{2} \and Terrence Chen\inst{1} \and Jian Xu\inst{2} \and Shanhui Sun\inst{1}}

\institute{United Imaging Intelligence, Cambridge, MA, USA \and UIH America, Inc., Houston, TX, USA \and United Imaging Healthcare, Shanghai, China}

\authorrunning{E. Chen et al.}
\titlerunning{ }

\maketitle

\begin{abstract}
Multi-contrast MRI images provide complementary contrast information about the characteristics of anatomical structures and are commonly used in clinical practice. Recently, a multi-flip-angle (FA) and multi-echo GRE method (MULTIPLEX MRI) has been developed to simultaneously acquire multiple parametric images with just one single scan. However, it poses two challenges for MULTIPLEX to be used in the 3D high-resolution setting: a relatively long scan time and the huge amount of 3D multi-contrast data for reconstruction. Currently, no DL based method has been proposed for 3D MULTIPLEX data reconstruction. We propose a deep learning framework for undersampled 3D MRI data reconstruction and apply it to MULTIPLEX MRI. The proposed deep learning method shows good performance in image quality and reconstruction time.
\end{abstract}



 

\section{Introduction}

Multi-contrast MRI images provide complementary contrast information about the characteristics of anatomical structures and are commonly used in clinical practice. Recently, a multi-flip-angle (FA) and multi-echo GRE method, hereinafter named as MULTIPLEX MRI, has been developed to simultaneously acquire multiple parametric images with just one single scan \citep{chen2018strategically,Ye2019swi}, which includes perfectly aligned PDW, T1W, T2* mapping and QSM images. However, it poses two challenges for MULTIPLEX to be used in the 3D high-resolution setting: a relatively long scan time and the huge amount of 3D multi-contrast data for reconstruction. To accelerate the 3D MULTIPLEX data acquisition, undersampled k-space data can be acquired. However, this makes the reconstruction of 3D MULTIPLEX data even more challenging. Since the compressed sensing (CS) algorithms are often iterative and time-consuming, it is difficult for CS-based method to reconstruct such a large amount of high-resolution 3D data within a clinically acceptable duration. On the other hand, deep learning (DL) methods only take one forward pass of the trained model to generate the reconstructed images and therefore such methods are much faster. Currently, no DL based method has been proposed for 3D MULTIPLEX data reconstruction. 

In this paper, we propose a DL framework for undersampled 3D MULTIPLEX MRI data reconstruction. The proposed method shows good performance in image quality and reconstruction time.


\section{Methods}
 
\textbf{Data collection}: The multi-FA and multi-echo GRE brain scans were performed on a 3T scanner (uMR 790 United Imaging Healthcare, Shanghai, China) with a 32-channel head coil, using the following parameters: FA1/FA2=4/16, 6$\sim$8 echoes with TE=2.1$\sim$20.8ms and TR=34.9ms, matrix size = $336\times278\times64$ with a voxel size of $0.69\times0.69\times2 mm^3$.  The data were acquired with parallel imaging, which was used as the ground truth to train and evaluate the DL model. The ground truth data were then retrospectively undersampled using the 3X and 5X Poisson-disk undersampling scheme in phase encoding and slice directions.

The data includes total 49 scans and we randomly split the data into 42 scans (662 3D echo data) for training and 5 scans (75 3D echo data) for testing.

\textbf{Reconstruction framework}:
With different echo and FA configures in one single scan, the 3D MULTIPLEX MRI can generate multiple (e.g., 7x2=14) 3D echo data, which is much larger than conventional 3D MRI data. To balance the reconstruction speed and GPU memory usage, we utilize the following strategy. 

We first used the DL model to construct all MULTIPLEX echo images.
We reconstructed each coil in each echo data separately to avoid loading the whole multi-echo multi-coil 3D data into the GPU memory. In the training stage, the readout (RO) was randomly cropped into 128 to further reduce memory consumption. In the testing stage, full RO was used. 

Then the DL reconstructed echo images were processed to generate the following parametric images: (1) Composite PDW and T1W: by averaging all the echo images from FA1 or FA2, respectively.
(2) T2* mapping: calculated using MDI method \citep{Ye2019mrr} on multi-echo images  
(3) QSM: the field map was first extracted from multi-echo images \citep{ye2019seed}, and was then used to generate QSM using the L2-norm optimization method with dynamic streaking artifact regularization \citep{Ye2019dynamic}. 

\textbf{DL model}: We employed a cascaded CNN model \citep{schlemper2017deep} with 3D convolutional layers and data consistency layers (Figure 1). The model is fully convolutional and thus can take input images with flexible sizes. The images were reconstructed using an Nvidia Tesla V100 GPU (16G memory).


\begin{figure}[htbp]
\centering
\includegraphics[width=\linewidth]{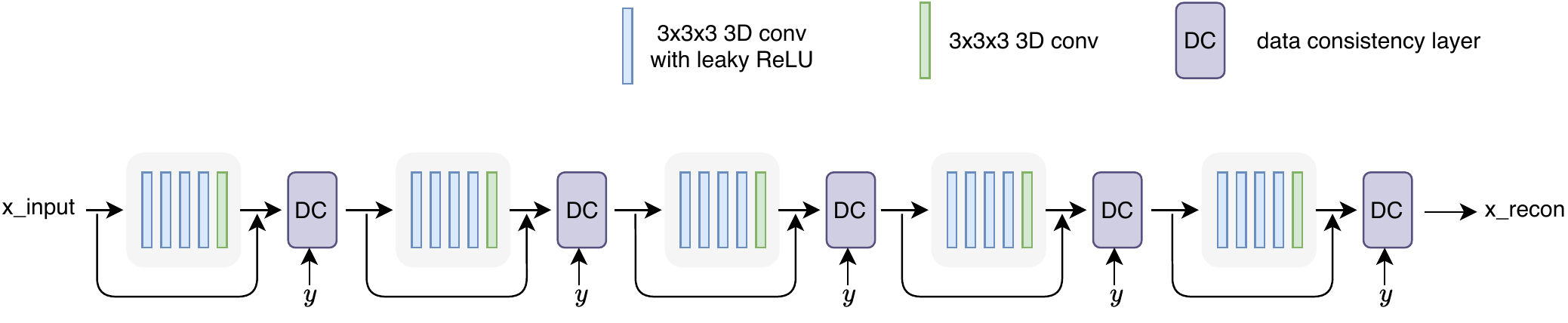}
\caption{The network architecture for 3D MULTIPLEX data reconstruction. The model includes five convolutional blocks and each block contains five 3D convolutional layers and one data consistency layer. The feature maps are 48 for the first four 3D convolutional layers and 2 for the last 3D convolutional layer in each block. The x, y indicate image and kspace, respectively. The real and imaginary numbers of complex values are transformed into two channels and fed into the network. }
\label{fig:network}
\end{figure}

\begin{figure}[htbp]
\centering
\includegraphics[width=\linewidth]{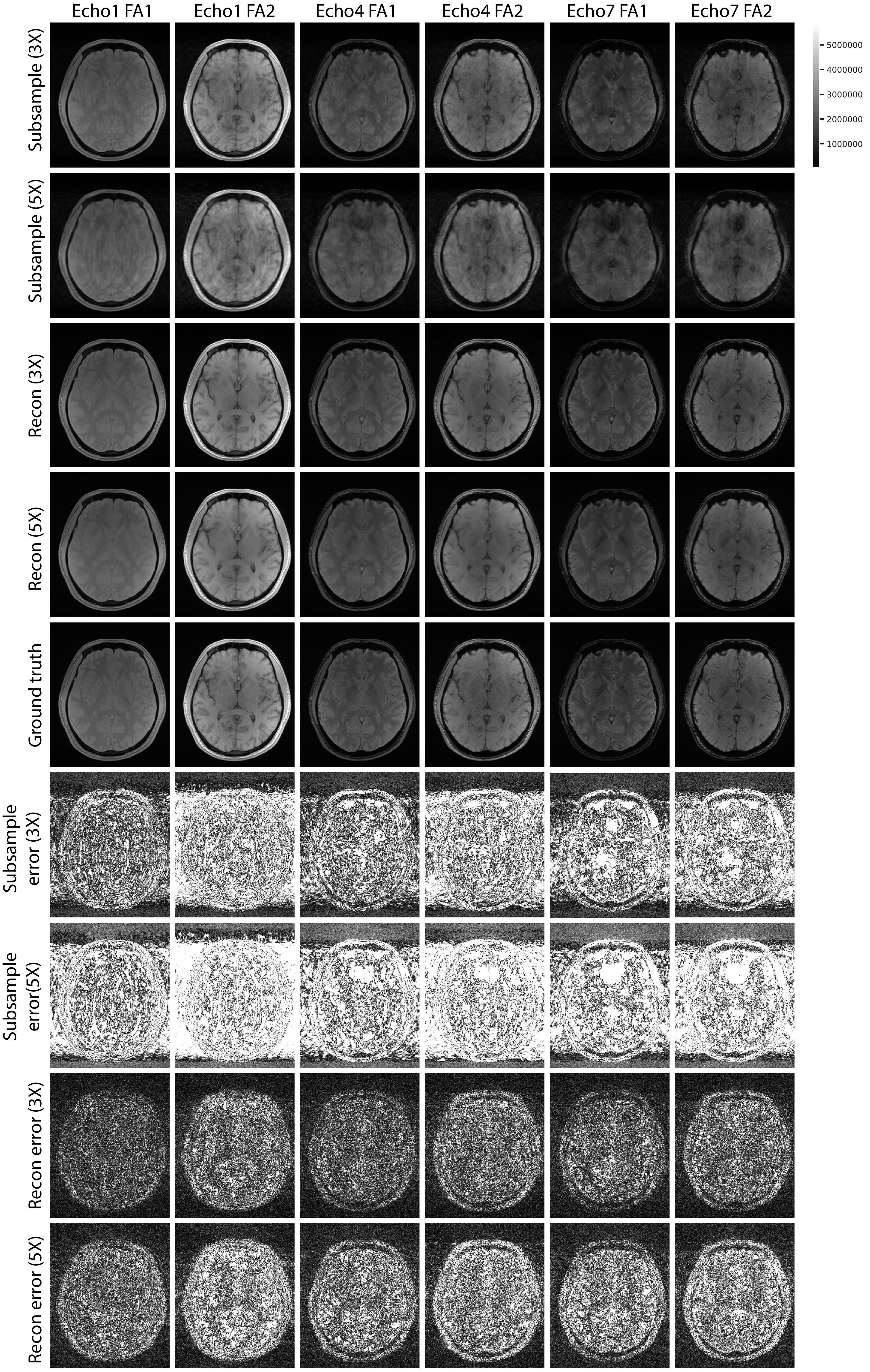}
\caption{Examples of reconstructed MULTIPLEX echo images by the proposed deep learning method at 3X and 5X accelerations. FA1 and FA2 indicate two different flip angle configurations. Three (Echo1, Echo4 and Echo7) out of seven echo configures are showed due to space constraints. All errors are multiplied by 50 for better visualization. The same 2D axial slice from each 3D image is plotted. 
}
\label{fig:echo_images}
\end{figure}

\begin{figure}[htbp]
\centering
\includegraphics[width=\linewidth]{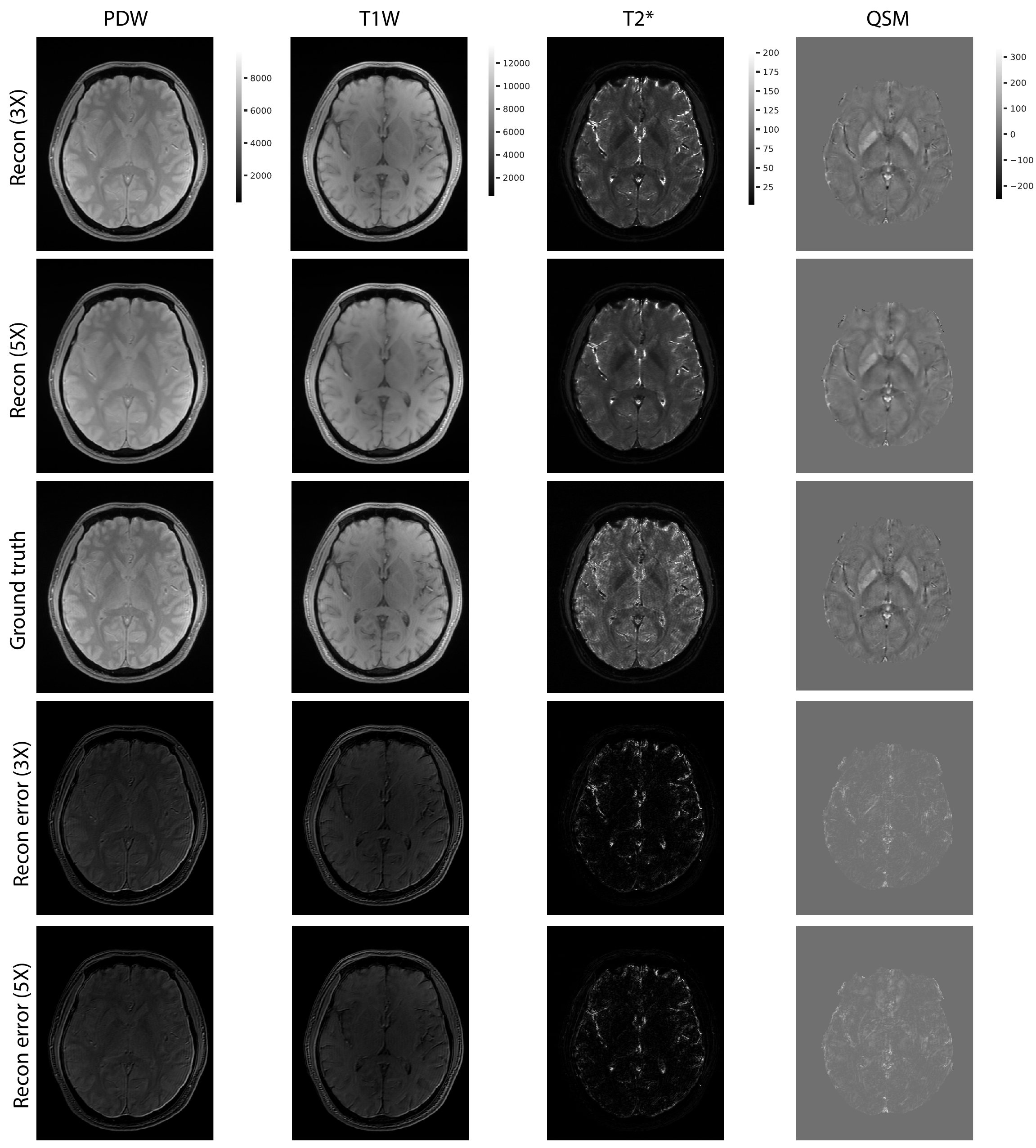}
\caption{Examples of reconstructed PDW, T1W, T2* and QSM images at 3X and 5X accelerations. The parametric images were calculated based on all the reconstructed echo images in the scan. The same 2D axial slice from each 3D image is plotted. }
\label{fig:map_image}
\end{figure}


\section{Results}
Figure 2 shows the examples of reconstructed MULTIPLEX echo images from the test dataset.  The reconstruction results at 3X acceleration are quite similar to the ground truth, while the results at 5X acceleration show relatively worse reconstruction quality. Compared to the ground truth, the DL reconstructions have less ringing artifacts and noise. Table 1 shows the quantitative results for the reconstructed echo images at 3X and 5X accelerations, which are consistent with the visual examples in Figure 2. Based on the reconstructed echo images, we also calculated the parametric images such as PDW, T1W, T2* and QSM (Figure 3 and Table 2), which are also similar to the ground truth results. 

For one 3D MULTIPLEX echo image, the DL model took 66 seconds to reconstruct all coil images with 15G GPU memory usage. We attempted to run a CS-based 3D reconstruction algorithm on the same data. However, it either ran out of memory using the GPU or took more than one hour to run using the CPU.      

\section{Discussion and Conclusion}
It is challenging to reconstruct the images from the large 3D MULTIPLEX data in a reasonably fast time. In this paper, we proposed to take advantage of the fast reconstruction speed of the DL model and reconstruct each echo image independently. The strategy leads to good reconstruction results and fast reconstruction time. Other DL model architectures for mobile devices such as MobileNet can be further explored to alleviate the computational burden, especially the GPU memory usage. With less GPU memory consumption, the multiple echo images can be fed into the model and the correlation between them can be learned by the model, which can potentially improve the reconstruction quality. 


\begin{table}[htbp] 
\caption{Quantitative results on the reconstructed 3D MULTIPLEX echo images. The mean and standard deviation of PSNR and SSIM are showed.
}
\centering
  \begin{tabular}{ccc}
    \toprule
    Acceleration & PSNR  & SSIM\\
    \midrule
    3X & 41.41 (2.63) &    0.97 (0.01)  \\ 
    5X & 36.86 (3.20) &    0.93 (0.03)  \\ 
    \bottomrule
  \end{tabular}
  \label{table:echo_image_measures}
\end{table}

\begin{table}[!htbp] 
\caption{Quantitative results on the PDW, T1W, T2* and QSM images. The mean and standard deviation of RMSE are showed. The parametric images have negative values and therefore PSNR and SSIM are not appropriate measures.
}
\centering
  \begin{tabular}{ccccc}
    \toprule
    Acceleration &  PDW &  T1W & T2* & QSM\\
    \midrule
    3X & 51.8 (5.1) & 52.7 (4.8) & 13.6 (3.6) & 9.3 (4.2)\\ 
    5X & 50.7 (6.7) & 54.6 (7.7) & 13.0 (3.5) & 11.2 (5.4)  \\ 
    \bottomrule
  \end{tabular}
  \label{table:echo_image_measures}
\end{table}


\bibliographystyle{abbrvnat}

\bibliography{references}
\end{document}